# Interpretation of the Omori law

**Anatol V. Guglielmi**

*Schmidt Institute of Physics of the Earth, Russian Academy of Sciences, ul. B. Gruzinskaya 10, Moscow, 123995 Russia*

*e-mail: guglielmi@mail.ru*

**Abstract**

The known Omori law is presented in the form of differential equation that describes the evolution of the aftershock activity. This equation is derived hypothetically with taking into account deactivation of the faults in epicentral zone of the main shock. A generalization of the Omori law is proposed.

1. **Introduction**

After the occurrence of the main shock, the source of the earthquake starts to relax. Speaking metaphorically, it gradually cools. However, this is much unlike the regular cooling of a heated homogeneous body. On the contrary, a source, which is a highly nonlinear structured system, undergoes complex processes manifesting themselves, in particular, by a sustained aftershock activity. In 1894 Fusakichi Omori [1] found that in average the aftershocks frequency decreases hyperbolically with time after a strong earthquake:

$$n(t) = \frac{k}{c+t}. \qquad (1)$$

Here $k > 0$, $c > 0$, $t \geq 0$. Formula (1) is called Omori law.

Nowadays the formula

$$n(t) = \frac{k}{(c+t)^p} \qquad (2)$$

is commonly used instead of (1). It has been proposed by Tokuji Utsu [2] (see also the review paper [3]). The parameter $p$ varies from place to place and from case to case. For example, $p$ varies from 0.5 to 1.5 with mean value $p = 1.08$ for aftershock sequences in California [4].

In this paper, we try to justify the Omori law in the form of (1), and we propose a generalization that is different from (2).

## 2. Reformulation

Firstly, we assume that the Omori law in the form of (1) is exact in some ideal sense. Secondly, we look at the formula (1) as the solution of a differential equation that describes the evolution of the aftershock activity. It is readily seen that the evolution equation has the form

$$\frac{dn}{dt} + \sigma n^2 = 0. \qquad (3)$$

Indeed, the solution of (3)

$$n(t) = n_0 (1 + \sigma n_0 t)^{-1} \qquad (4)$$

coincides with (1), if we put $\sigma = k^{-1}$, $n_0 = k/c$.

At first glance, it seems that equation (3) is not more than another form of the hyperbolic Omori law, but only at first. Let us take into account that the evolution equations are very useful in the analysis of natural phenomena. A look at the Omori law as the differential equation (3) has an advantage that opens up a new possibility for physical interpretation of the aftershock sequences.

## 3. Interpretation

Equation (3) encourages us to see some similarity between the attenuation of aftershock activity, and a decrease in the ionospheric plasma density due to recombination of charges of the opposite signs. Recall that the radiative recombination of pairs of the oppositely charged particles occurs according to the scheme:

$$O_2^+ + e^- \to O_2 + \hbar\omega. \tag{5}$$

Here $O_2$ is the oxygen molecule, $O_2^+$ is the oxygen ion, $e^-$ is the electron, $\hbar\omega$ is the photon. As a result, the pair of charges disappears, and the neutral molecule and photon appear. Let $n_+$ ($n_-$) is the density of positive (negative) charges, and $n = (n_+ + n_-)/2$. Then the equation of recombination is

$$dn/dt = -\sigma n_+ n_-, \tag{6}$$

where $\sigma$ is the coefficient of recombination. In the ionosphere, we have $n_+ \approx n_-$ with a high degree of accuracy. Hence (6) is practically coincides with (3).

Let us try to deepen this formal analogy. To do this, we need to find "a pair" in the earth's crust, similar to the pair of oppositely charged particles in the ionosphere. At this point of our reasoning the idea of a pair of adjacent sides of fault arises. Let us denote by symbol ↑↓ the fault to which the shear stress is applied. Such fault will be called active. In active fault the rupture can happen with some probability and then an earthquake occurs. By analogy with the reaction (5), we write symbolically

$$\uparrow + \downarrow \to \| + aftershock. \tag{7}$$

Here $\|$ is the passive fault (without shear stress). Let $n$ is the number of active faults in the epicentral region of the main shock. By analogy with (6), we write

$$dn/dt = -\sigma n_\uparrow n_\downarrow. \tag{8}$$

This implies (3) since $n_\uparrow = n_\downarrow$. The value $\sigma$ in (8) is the mean frequency of deactivation of the active faults.

### 4. Discussion

It is reasonable to assume that $\sigma$ in (3) is a function of time in the cooling source of the main shock. If so, then instead of (4) we have the following solution of the equations (3):

$$n(t) = n_0 \left[ 1 + n_0 \int_0^t \sigma(t') dt' \right]^{-1} \tag{9}$$

Thus, the observed deviations from a strictly hyperbolic Omori law (1) may be associated with the nonstationarity of the earthquake source.

Another type of deviations from the Omori law (1) arises due to the phenomenon of round-the-world seismic echo. We will not dwell on this, referring the interested reader to the papers [5, 6].

### 5. Conclusion

It is shown that the Omori law may be presented in the form of differential equation that describes the evolution of the aftershock activity. This equation is interpreted hypothetically with taking into account deactivation of the faults in epicentral zone of the main shock. The general solution of the evolution equation makes it possible to take into account a nonstationarity of the earthquake source.

*Acknowledgments.* I am grateful to A.L. Buchachenko, B.I. Klain, S.M. Molodenskii, A.S. Potapov, L.E. Sobisevich, A.L. Sobisevich, A.D. Zavyalov, and O.D. Zotov for their useful discussions on physics of the earthquakes. The key results of this study were outlined in my presentation at the session of the Scientific Board of the Schmidt Institute of Physics of the Earth of the Russian Academy of Sciences (RAS) on April 13, 2016. I thank the participants of the session for their interest and, particularly, for their critical comments. The paper under the title "Interpretation of the Omori law" accepted for publication in the journal "Izvestiya, Physics of the Solid Earth", 2016, No. 5. The work was supported by basic research program 15 of the Presidium of RAS and by the Russian Foundation for Basic Research under project No. 15-05-00491.**References**

[1] Omori, F., On the aftershocks of earthquake // J. Coll. Sci. Imp. Univ. Tokyo, 1894, V. 7, P. 111–200.

[2] Utsu, T., A statistical study on the occurrence of aftershocks // Geophys. Mag., 1961, V. 30, P. 521–605.

[3] Utsu, Tokuji, Yosihiko Ogata, and Ritsuko S. Matsu'ura, The centenary of the Omori formula for a decay law of aftershock activity // J. Phys. Earth, 1995, V. 43, P. 1-33.

[4] Reasenberg, Paul A., and L.M. Jones, Earthquake hazard after a mainshock in California // Science, 1989, V. 243, P. 1173 - 1176.

[5] Guglielmi, A.V., Foreshocks and aftershocks of strong earthquakes in the light of catastrophe theory // Physics-Uspekhi, 2015, V. 58, No.4, P. 384–397.

[6] Guglielmi, A.V., On self-excited oscillations of the Earth // Izvestiya, Physics of the Solid Earth, 2015, V. 51, No. 6, P. 920–923. DOI: 10.1134/S1069351315040011
5